# Quantum spectral thermodynamics and active learning enable million-scale exploration of high-entropy ceramics

Jie Sun[1], Yiheng Shen[2]

1. Chimie ParisTech, PSL University, CNRS, Institute of Chemistry for Life and Health Sciences, Chemical Theory and Modelling Group, F-75005 Paris, France.

2. Materials Genome Institute, Shanghai University, Shanghai, China.

## Abstract

Understanding phase stability and navigating vast compositional spaces in multicomponent solids remain central challenges in solid-state chemistry. Here, we develop a quantum spectral thermodynamic framework connecting interaction-induced phonon spectral broadening to free energy, alongside an uncertainty-guided active-learning workflow that explores 7.7 million high-entropy ceramic configurations at density-functional-theory fidelity, achieving a $10^5$-fold acceleration. We show that phonon self-energy effects arising from chemical disorder provide an intrinsic vibrational contribution to thermodynamic stabilization beyond ideal configurational entropy, compensating unfavorable mixing enthalpies and suppressing phase separation. Across the chemical space, we uncover a robust ∼12 at.% solute threshold separating strengthening and softening regimes, associated with the filling of metal-carbon antibonding states. Chemical disorder further enables an unusual combination of high-temperature mechanical stiffness and low thermal conductivity, together with anomalous temperature-dependent lattice heat transport. This work establishes a quantum spectral foundation for connecting many-body interactions to thermodynamics and phase stability, while providing a scalable framework for exploring previously inaccessible multicomponent chemical spaces.

## Introduction

For nearly a century, the design of inorganic crystals has largely been guided by narrow solubility limits and well-defined stoichiometries, with functional materials predominantly explored within conventional binary and ternary compositional spaces[1,2]. The emergence of the high-entropy concept two decades ago challenged this paradigm[3,4], enabling multiple principal elements to coexist on a common lattice and opening vast compositional spaces governed by chemical disorder[5]. In these multicomponent crystals, atomic-size mismatch and heterogeneous electronic structures generate pronounced local distortions and fluctuations in bonding, giving rise to properties that depart markedly from simple rule-of-mixtures expectations[6–9]. Yet, a fundamental challenge remains: long-range structural order can coexist with severe local chemical disorder[10,11], raising the question of how such microscopic heterogeneity governs macroscopic properties and phase stability.

In solid-state thermodynamics, the stabilization of single-phase solid solutions is traditionally evaluated through the Gibbs free energy of mixing, $\Delta G_{\mathrm{mix}} = \Delta H_{\mathrm{mix}} - T\Delta S_{\mathrm{mix}}$, which reflects the competition between the enthalpic cost of chemical mixing and the entropic stabilization of disorder[12,13]. The ideal configurational contribution, $-k_{\mathrm{B}}T\sum c_i \ln c_i$, is readily evaluated, whereas excess contributions arising from multicomponent interactions become increasingly difficult to capture as chemical complexity grows. Semi-empirical approaches such as CALPHAD address this challenge through effective parameters[14–16], but do not explicitly resolve the microscopic quantum interactions underlying the free-energy landscape. First-principles treatments have incorporated vibrational contributions through temperature-dependent phonon renormalization[17,18], yet generally retain a quasiparticle description without explicitly resolving the many-body spectral effects that shape finite-temperature vibrational thermodynamics. In multicomponent solids, local bonding disorder and strong chemical anharmonicity provide a natural source of such interactions, raising the question of whether their spectral effects can generate additional vibrational stabilization sufficient to compensate unfavorable mixing enthalpies and suppress phase separation. Meanwhile, direct density functional theory (DFT) calculations and experimental exploration of these vast compositional spaces remain prohibitively expensive[19], creating a severe bottleneck in high-quality training data. Although universal machine-learning potentials have begun to alleviate this bottleneck by approaching DFT accuracy across broad chemical domains[20–23], their computational cost can still limit large-scale dynamical simulations[24,25]. Consequently, both the microscopic origin of finite-temperature stabilization and the exploration of its consequences across vast multicomponent chemical spaces remain challenging.

Here, we develop a quantum spectral thermodynamic framework for describing finite-temperature phase stability in multicomponent solids, using high-entropy ceramic carbides as a model system. We show that chemical bond fluctuations generate phonon self-energy effects whose spectral broadening provides an additional thermodynamic driving force beyond ideal configurational entropy, lowering the vibrational free energy and contributing to the stabilization of single-phase solid solutions. To explore the resulting compositional landscape, we develop an uncertainty-guided active-learning workflow that combines a universal machine-learning potential with surrogate models to map the mechanical-property landscape at million-configuration scale. This approach reveals a robust solute concentration threshold separating strengthening and softening regimes and enables us to trace the microscopic origin of shear softening to the progressive filling of metal–carbon antibonding states. Furthermore, a lightweight model derived from the high-fidelity potential enables extended high-temperature simulations, revealing how chemical disorder can simultaneously sustain mechanical stiffness and modify lattice heat transport at elevated temperatures.

## Results and discussions

### Quantum many-body effects of multi-element materials

In chemically disordered high-entropy lattices, the configuration-averaged vibrational eigenstates are projected onto an effective medium virtual lattice indexed by $(q, \lambda)$. Starting from atomic displacements, the total lattice Hamiltonian is partitioned into an unperturbed harmonic reference $\hat{H}_0$ and a cubic perturbation, transformed into second-quantized form via field operators $\hat{A}_{q\lambda} = \hat{a}_{q\lambda} + \hat{a}^{\dagger}_{-q\lambda}$:

$$\hat{H} = \hat{H}_0 + \sum_{q_1\lambda_1, q_2\lambda_2, q_3\lambda_3} V^{(3)}\,(q_1\lambda_1, q_2\lambda_2, q_3\lambda_3)\hat{A}_{q_1\lambda_1}\hat{A}_{q_2\lambda_2}\hat{A}_{q_3\lambda_3}, \quad (1)$$

where $V^{(3)}$ denotes the projected third-order scattering vertex (Supplementary Information (SI) section 1). At finite temperatures ($T > 0$), microscopic interactions are governed by the dressed bosonic Green's function $D(q\lambda, i\omega_n)$, which satisfies the exact Dyson equation (SI section 2):

$$\mathcal{D}(q\lambda, i\omega_n) = \frac{2\omega_{q\lambda}}{(i\omega_n)^2 - \omega_{q\lambda}^2 - 2\omega_{q\lambda}\Sigma(q\lambda, i\omega_n)} \quad (2)$$

The self-energy $\Sigma(q\lambda, i\omega_n)$ is evaluated via the one-loop Feynman bubble diagram (Fig. 1a), where the incoming phonon $(q\lambda)$ virtually decays into two intermediate modes $(q_1\lambda_1)$ and $(q_2\lambda_2)$ through the cubic vertex $V^{(3)}$. By performing Matsubara frequency summations over internal poles followed by analytic continuation to the real frequency axis ($i\omega_n \rightarrow \omega + i0^+$) via the Dirac-Plemelj relation $\lim_{\eta\rightarrow 0^+} \frac{1}{x \pm i\eta} = \mathcal{P}(\frac{1}{x}) \mp i\pi\delta(x)$, the retarded self-energy is resolved into a real frequency shift $\Delta(q\lambda, \omega)$ and an imaginary linewidth damping $\Gamma(q\lambda, \omega) \equiv -\mathrm{Im}\Sigma(q\lambda, \omega + i0^+)$ (SI section 3):

$$\Gamma(q\lambda, \omega) = \frac{\pi}{18\hbar^2}\sum_{1,2} V^{(3)}\,|(-q\lambda, q_1\lambda_1, q_2\lambda_2)|^2[(n_1 + n_2 + 1)\delta(\omega - \omega_1 - \omega_2) + (n_2 - n_1)\delta(\omega + \omega_1 - \omega_2)], \quad (3)$$

where $n_i = [\exp(\beta\hbar\omega_i) - 1]^{-1}$ represents the Bose-Einstein occupancy. This microscopic damping dynamically regularizes the sharp harmonic delta function into a continuous, phonon linewidth spectral function $A(q\lambda, \omega) = -\frac{1}{\pi}\mathrm{Im}D^R(q\lambda, \omega)$:

$$A(q\lambda, \omega) = \frac{1}{\pi}\frac{4\omega_{q\lambda}^2\Gamma(q\lambda, \omega)}{[\omega^2 - \omega_{q\lambda}^2 - 2\omega_{q\lambda}\Delta(q\lambda, \omega)]^2 + [2\omega_{q\lambda}\Gamma(q\lambda, \omega)]^2} \quad (4)$$

Because the coordinate–coordinate phonon spectral function obeys a frequency-weighted sum rule rather than a unit-normalization condition, we construct a normalized positive-frequency effective spectral weight $\rho_{\mathbf{q}\lambda}(\omega, T) \equiv A(\mathbf{q}\lambda, \omega, T)/\int_0^\infty A(\mathbf{q}\lambda, \omega', T)\mathrm{d}\omega'$, which satisfies $\int_0^\infty \rho_{\mathbf{q}\lambda}(\omega, T)\mathrm{d}\omega = 1$ by construction (SI section 4). Within a practical spectral quasiparticle approximation motivated by the stationary Luttinger-Ward functional, the effective vibrational Helmholtz free energy $F_{\mathrm{vib}}^{\mathrm{spec}}(T)$ is evaluated by performing a spectral average of the single-mode free-energy kernel:

$$F_{\mathrm{vib}}^{\mathrm{spec}}(T) = k_{\mathrm{B}}T\sum_{\mathbf{q}\lambda}\int_0^\infty \mathrm{d}\,\omega\rho_{\mathbf{q}\lambda}(\omega, T)\ln\left[2\sinh\left(\frac{\hbar\omega}{2k_{\mathrm{B}}T}\right)\right] \tag{5}$$

Under the standard fixed-spectrum approximation ($\partial\rho/\partial T \simeq 0$; see SI section 4 for complete derivatives), the corresponding vibrational internal energy and entropy are obtained by integrating their respective oscillator kernels:

$$U_{\mathrm{vib}}^{\mathrm{spec}}(T) \simeq \sum_{\mathbf{q}\lambda}\int_0^\infty \mathrm{d}\,\omega\rho_{\mathbf{q}\lambda}(\omega, T)\left[\frac{\hbar\omega}{2}\coth\left(\frac{\hbar\omega}{2k_{\mathrm{B}}T}\right)\right] \tag{6}$$

$$S_{\mathrm{vib}}^{\mathrm{spec}}(T) \simeq k_{\mathrm{B}}\sum_{\mathbf{q}\lambda}\int_0^\infty \mathrm{d}\,\omega\rho_{\mathbf{q}\lambda}(\omega, T)\left[\frac{\hbar\omega}{2k_{\mathrm{B}}T}\coth\left(\frac{\hbar\omega}{2k_{\mathrm{B}}T}\right) - \ln\left(2\sinh\frac{\hbar\omega}{2k_{\mathrm{B}}T}\right)\right] \tag{7}$$

The macroscopic driving force governing phase stability against decomposed reference phases is then captured by the free energy of mixing: $\Delta G_{\mathrm{mix}}(T) \simeq \Delta E_{\mathrm{static}} - TS_{\mathrm{conf}} + \Delta F_{\mathrm{vib}}^{\mathrm{har}}(T) + \Delta F_{\mathrm{vib}}^{\mathrm{excess}}(T)$, where $\Delta F_{\mathrm{vib}}^{\mathrm{har}}(T) = F_{\mathrm{vib}}^{\mathrm{har}} - \sum_i x_i F_{\mathrm{vib}}^{i,\mathrm{har}}$ denotes the baseline harmonic vibrational contrast, and $\Delta F_{\mathrm{vib}}^{\mathrm{excess}}(T) = [F_{\mathrm{vib}}^{\mathrm{spec}}(T) - F_{\mathrm{vib}}^{\mathrm{har}}(T)] - \sum_i x_i [F_{\mathrm{vib}}^{i,\mathrm{spec}}(T) - F_{\mathrm{vib}}^{i,\mathrm{har}}(T)]$ consolidates the finite-temperature quantum many-body stabilization beyond the harmonic baseline. Because the bosonic free-energy kernel is concave in frequency, Jensen's inequality implies that a finite spectral width lowers the spectrally averaged free energy relative to the kernel evaluated at the corresponding spectral mean frequency. When the spectral mean is conserved relative to the harmonic reference, this yields a non-positive excess contribution, $\Delta F_{\mathrm{vib}}^{\mathrm{excess}} \leq 0$, with equality only in the zero-width limit (SI, Section 5).

We applied the framework to two high-entropy carbides representing light- and heavy-element systems: the lighter $(Al_{0.2}Sc_{0.2}Mn_{0.2}Ti_{0.2}V_{0.2})C$ and the heavier $(Hf_{0.2}Cr_{0.2}Ta_{0.2}Ti_{0.2}Zr_{0.2})C$ (Fig. 1b,c). Both systems were constructed as rock-salt special quasi-random structures (SQS) to represent chemical disorder[26]. To reduce the computational cost of calculating second- and third-order interatomic force constants in large supercells, structural relaxations and force evaluations were performed using the universal machine-learning potential MACE[21]. Harmonic phonon dispersions confirm the dynamical stability of both systems (Fig. 1d,e). Within the one-loop self-energy treatment, the phonon linewidth $\Gamma(q\lambda, \omega)$ increases monotonically with temperature from 300 to 1500 K in both carbides (Fig. 1f,g). Details of the SQS construction, MACE validation against DFT lattice constants, and spectral-broadening calculations are provided in SI, Sections 6-9.

The thermodynamic consequence of finite phonon linewidths is reflected in the temperature dependence of the excess vibrational free energy (Fig. 1h). Here, the self-energy contains only intrinsic phonon-phonon interactions. From 300 to 1500 K, $\Delta F_{\mathrm{vib}}^{\mathrm{excess}}$ decreases monotonically from nearly zero to -15.5 meV/atom for $(Al_{0.2}Sc_{0.2}Mn_{0.2}Ti_{0.2}V_{0.2})C$ and -19.2 meV/atom for $(Hf_{0.2}Cr_{0.2}Ta_{0.2}Ti_{0.2}Zr_{0.2})C$. Despite their distinct phonon dispersions, both systems therefore exhibit an increasingly negative vibrational free-energy contribution with temperature. Decomposition into vibrational enthalpy and entropy shows that the enthalpic contribution remains below 1 meV/atom over the entire temperature range, leaving the free-energy change dominated by the entropic term (Fig. 1i). At 1500 K, $-T\Delta S_{\mathrm{vib}}$ reaches -16.0 and -19.5 meV/atom for the lighter and heavier carbides, respectively, corresponding to approximately 15-20% of the ideal configurational entropy contribution ($-T\Delta S_{\mathrm{config}} = -0.1$ eV/atom). The linewidths remain well below the corresponding phonon frequencies, ($\Gamma_{\mathbf{q}\lambda} \ll \omega_{\mathbf{q}\lambda}$), throughout this temperature range (SI Section 10 and Fig. S1), supporting the validity of the perturbative treatment. These results establish the intrinsic phonon-phonon contribution as a baseline for the subsequent inclusion of mass-disorder broadening.

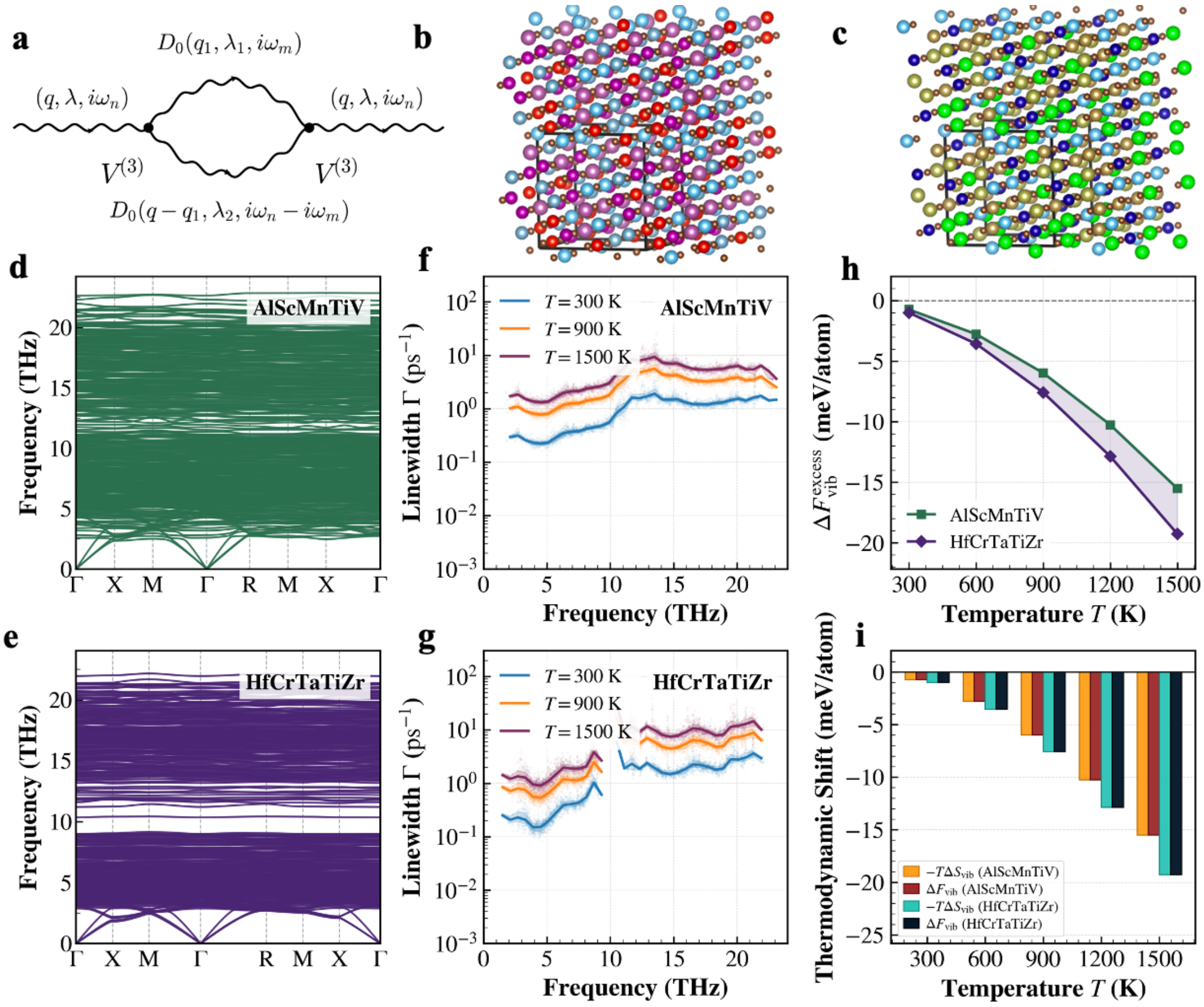

**Figure 1. Many-body vibrational renormalization and thermodynamic phase stabilization. a** One-loop Feynman bubble diagram of the phonon self-energy driven by third-order anharmonic scattering vertices $V^{(3)}$. Supercell configurations of **b** $(Al_{0.2}Sc_{0.2}Mn_{0.2}Ti_{0.2}V_{0.2})C$ and **c** $(Hf_{0.2}Cr_{0.2}Ta_{0.2}Ti_{0.2}Zr_{0.2})C$ systems. Harmonic phonon dispersion relations at 0 K for **d** $(Al_{0.2}Sc_{0.2}Mn_{0.2}Ti_{0.2}V_{0.2})C$ and **e** $(Hf_{0.2}Cr_{0.2}Ta_{0.2}Ti_{0.2}Zr_{0.2})C$. **f, g** Frequency-dependent microscopic phonon linewidths $\Gamma(q\lambda,\omega)$ at 300 K, 900 K, and 1500 K, where dots indicate discrete modes and solid lines denote binned median envelopes. **h** Net excess vibrational free energy change as a function of temperature, with the shaded region highlighting the stabilization gap between the two systems. **i** Thermodynamic decomposition of the spectral shift into vibrational entropy ($-T\Delta S_{\mathrm{vib}}$) and net free energy ($\Delta F_{\mathrm{vib}}$) contributions across temperatures. Chemical formulas omit carbon and stoichiometric ratios in panels for brevity.

Having established that interaction-induced spectral broadening lowers the vibrational free energy of the single-phase lattice, we next examine its effect on phase stability against decomposition, using $(Hf_{0.2}Cr_{0.2}Ta_{0.2}Ti_{0.2}Zr_{0.2})C$ as a representative system. Although equimolar high-entropy carbides can decompose into non-equimolar phases, we benchmark the free energy against the corresponding terminal carbides and graphite. These reference phases define the relevant 0 K convex-hull boundary, providing a stringent reference for assessing finite-temperature stabilization of the high-entropy phase. Hf, Ta, Ti and Zr form stoichiometric rock-salt monocarbides, whereas Cr preferentially forms the carbon-deficient orthorhombic $Cr_3C_2$ phase. Stoichiometric decomposition therefore requires graphite precipitation:

$$(\mathrm{Hf_{0.2}Cr_{0.2}Ta_{0.2}Ti_{0.2}Zr_{0.2}})\mathrm{C} \longrightarrow 0.2\mathrm{HfC} + 0.2\mathrm{TaC} + 0.2\mathrm{TiC} + 0.2\mathrm{ZrC} + \frac{1}{15}\mathrm{Cr_3C_2} + \frac{1}{15}\mathrm{C_{graphite}}$$

The optimized structures and lattice parameters are provided in SI, Section 11 and Table S2. Harmonic phonon dispersions for all reference phases are shown in Fig. 2a-f. The absence of imaginary frequencies confirms their dynamical stability and provides a basis for evaluating their harmonic vibrational free energies.

Fig. 2g compares the phonon linewidths of the high-entropy carbide and its decomposition products at 300 K. Graphite exhibits the weakest damping across the vibrational spectrum, consistent with its strong $sp^2$-bonded lattice. Fig. 2h summarizes the temperature-dependent mixing Gibbs free energy across four levels of description: static distortion with ideal configurational entropy (Tier 0), harmonic phonons (Tier 1), intrinsic phonon-phonon interactions (Tier 2a), and the combined intrinsic and mass-disorder contributions (Tier 2b). The Tier 1 result closely follows the Tier 0 baseline, with the harmonic vibrational free-energy difference remaining below 2 meV/atom throughout the temperature range (Fig. 2i). Thus, harmonic vibrations provide only a negligible contribution to the phase-stability balance. Within this approximation, $\Delta G_{\mathrm{mix}} = 0$ is reached at approximately 1140 K. Incorporating intrinsic phonon-phonon interactions lowers $\Delta F_{\mathrm{vib}}^{\mathrm{excess}}$ by approximately 7 meV/atom at 1500 K and shifts the corresponding phase boundary by about 53 K. The inclusion of mass disorder (SI, section 9) produces a substantially larger additional contribution: the excess vibrational free-energy gain reaches approximately -30 meV/atom at 1500 K, lowering the transition temperature to $\sim$890 K and giving a boundary shift of $\Delta T \approx$ -248 K.

The mass-disorder contribution enters the same spectral thermodynamic framework through the imaginary part of the phonon self-energy, without requiring any modification of the underlying thermodynamic formulation. In the present system, mass disorder substantially enhances the linewidth, particularly for high-frequency optical modes (SI, Section 12 and Fig. S2), and therefore provides the dominant additional source of spectral broadening. The perturbation Hamiltonian considered here contains lattice degrees of freedom, including cubic anharmonicity and mass disorder. Other dissipation channels, such as electron-phonon or spin-phonon interactions, can in principle contribute additively to the phonon self-energy within the same framework or roughly through Matthiessen's rule[27]. For the refractory carbide studied here, we restrict the calculation to intrinsic phonon-phonon interactions and mass disorder.

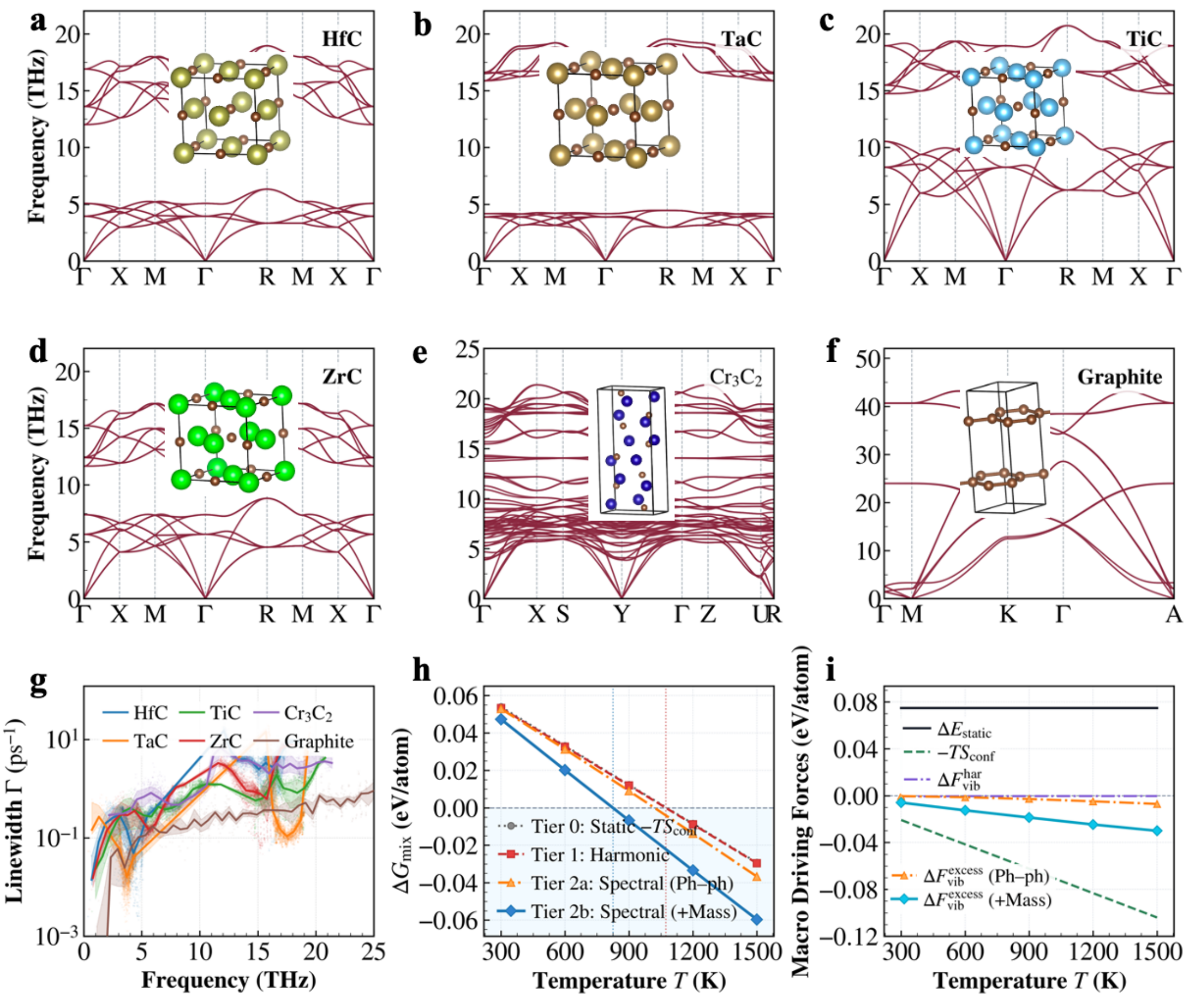

**Figure 2 | Phonon dispersion, linewidths, and phase stability of the high-entropy ceramic. a-f** Calculated phonon dispersions for the six reference phases (HfC, TaC, TiC, ZrC, $Cr_3C_2$, and graphite). **g** Phonon linewidths (Γ) as a function of frequency at 300 K; solid curves and shaded areas show the binned trends and data distributions. **h** Mixing Gibbs free energy ($\Delta G_{\text{mix}}$) as a function of temperature across four levels of description: static lattice distortion with configurational entropy (Tier 0), harmonic phonons (Tier 1), intrinsic phonon-phonon interactions (Tier 2a), and full dissipation incorporating chemical mass disorder (Tier 2b). The shaded region indicates single-phase stability ($\Delta G_{\text{mix}} < 0$). **i** Macroscopic contributions ($\Delta E_{\text{static}}, -TS_{\text{conf}}$ , and

$\Delta F_{\mathrm{vib}}^{\mathrm{har}}$) plotted alongside the net excess vibrational free energy ($\Delta F_{\mathrm{vib}}^{\mathrm{excess}}$) from intrinsic phonon-phonon interactions alone and with the inclusion of mass disorder.

**High-Throughput Computational Workflow**

With the microscopic picture of thermodynamic stabilization established, we now turn to the complementary challenge of exploring the vast compositional space of high-entropy ceramics. We developed an active-learning workflow to map the mechanical-property landscape of multicomponent carbides (Fig. 3), moving beyond isolated property extrema to capture a continuous range of mechanical responses relevant to diverse engineering requirements, from high stiffness to moderate-modulus compatibility.

We considered 14 carbide-forming metals (Sc, Ti, V, Cr, Mn, Al, Y, Zr, Nb, Mo, Hf, Ta, W and Ir) on the rock-salt lattice[28]. The initial database comprised 2,002 equimolar ($\mathrm{M_1^{0.2}M_2^{0.2}M_3^{0.2}M_4^{0.2}M_5^{0.2}}$)C configurations, together with 10,010 configurations centered on a single dominant metal ($\mathrm{M_1^{0.6}M_2^{0.1}M_3^{0.1}M_4^{0.1}M_5^{0.1}}$)C, and 20,020 centered on two equally dominant metals, ($\mathrm{M_1^{0.35}M_2^{0.35}M_3^{0.1}M_4^{0.1}M_5^{0.1}}$)C (Fig. 3, top). Within an 80-atom supercell containing 40 metal sites, a 5% compositional resolution was used to discretize the chemical space into 84 unique stoichiometric SQS templates (see SI Section 6 for structural details)[26]. These templates were then used by the committee ensemble to screen more than $7.7 \times 10^6$ candidate configurations across the non-equimolar compositional space. By minimizing short-range order within the finite supercell, the SQS construction provides consistent structural representations for evaluating elastic constants and mechanical properties.

To efficiently evaluate the elastic response of the disordered structures, we implemented an automated workflow based on the universal MACE interatomic potential[21]. For the equimolar ($\mathrm{Ti_{0.2}V_{0.2}W_{0.2}Ta_{0.2}Hf_{0.2}}$)C configuration with the highest Young's modulus, MACE reproduced the DFT equilibrium lattice dimensions and elastic moduli within 0.04% and 4%, respectively (SI, Section 13, Fig. S3, and Tables S3 and S4). Elastic constants were obtained from volume-conserving and hydrostatic strains applied to the relaxed SQS cells using the quasi-cubic rock-salt approximation, followed by Voigt-Reuss-Hill[29] averaging for the isotropic bulk, shear and Young's moduli (see SI section 14)[30]. The corresponding strain-energy calculations required over 7,000 $s$ with DFT on 64 CPU cores, compared with approximately 5 min using MACE. For large-scale screening, three ML models formed an uncertainty-guided committee based on the variance of predicted $E$, $\sigma^2 = \mathrm{Var}(\mathbf{F}_i)$. High-uncertainty structures were selected for additional MACE evaluation, while confident predictions were retained in the database. For representative high-modulus candidates ($E$ > 500 GPa), Monte Carlo atomic rattling was further used to generate thermally displaced structures, from which composition-specific ML potentials were constructed for finite-temperature tensile MD and interatomic force-constant calculations for lattice thermal transport[31].

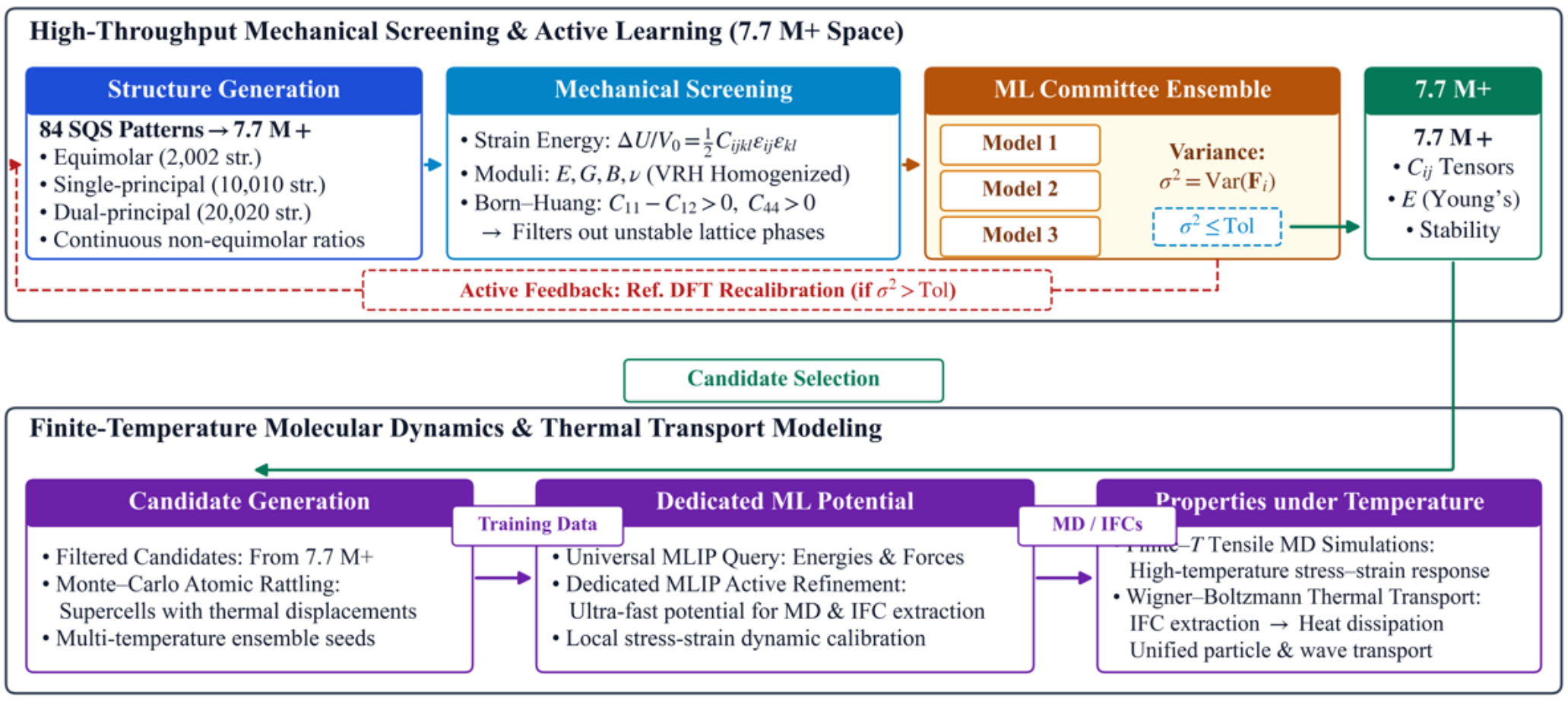

**Figure 3.** Overview of the active-learning framework for high-entropy carbide ceramics. The workflow integrates large-scale structure generation based on SQS patterns, rapid mechanical filtering using universal machine-learning potentials, uncertainty-driven active-learning refinement, and finite-temperature property predictions via molecular dynamics and Wigner-Boltzmann transport theory.

### Active-Learning Driven Mechanical Mapping across 7.7 Million Compositions

Using MACE to evaluate their mechanical response, we first tested the workflow on equimolar quinary compositions, $(\mathrm{M}_1^{0.2}\mathrm{M}_2^{0.2}\mathrm{M}_3^{0.2}\mathrm{M}_4^{0.2}\mathrm{M}_5^{0.2})\mathrm{C}$. After Born-Huang stability filtering, 1,967 of the 2,002 configurations were found to be mechanically stable. The resulting bulk, Young's and shear moduli span broad distributions, with mean values of 220.6, 336.8 and 135.6 GPa, respectively (Fig. 4a-c). We then trained a Crystal Graph Convolutional Neural Network (CGCNN) on this benchmark dataset to establish the structure-modulus mapping[32]. Five-fold cross-validation yielded an overall $R^2 = 0.98$, MAE of 7.95 GPa and RMSE of 14.12 GPa for Young's modulus (Fig. 4d). The training and validation losses decreased consistently without overfitting, with the five individual models achieving validation $R^2$ values tightly distributed between 0.94 and 0.98 (Fig. 4e,f).

With the initial training pool established from the equimolar, single-dominant and dual-dominant configurations, we deployed an active-learning framework with an ensemble of three CGCNN models to screen more than $7.7 \times 10^6$ non-equimolar candidate configurations (Fig. 4g,h). Configurations with an ensemble standard deviation of $\sigma > 20$ GPa were selected for MACE evaluation and added to the training set through iterative updates. The number of high-uncertainty configurations decreased from $3.68 \times 10^5$ in Loop 1 to $2.61 \times 10^5$, $2.14 \times 10^5$, and $1.76 \times 10^5$ in Loops 2–4, corresponding to an overall reduction of 52.1% (Fig. 4h). The remaining configurations were then evaluated directly with MACE. After incorporating the non-equimolar configurations, the validation MAE increased modestly to 11.90 GPa, while the committee maintained $R^2 > 0.95$ across the expanded chemical space (Fig. 4g). The resulting model was then used to predict Young's modulus for the full 7.7-million-configuration space (Fig. 4i).

Across the 7.7-million-configuration space, 33.20% ($2.56 \times 10^6$) of the structures have Young's moduli of 300-400 GPa, while 39.04% ($3.0 \times 10^6$) fall within 400-500 GPa. Ultra-stiff configurations with $E > 500$ GPa account for 15.51% ($1.20 \times 10^6$) of the space (Fig. 4i). We further validated the surrogate by comparing CGCNN predictions for representative stiff configurations with MACE calculations based on strain-energy fitting (Fig. 4j). The predicted-to-calculated ratios are centered near unity, with a MAPE of 0.42% and more than 95% of the structures within 0.5% error. With an inference time of approximately 0.1 *s* per structure, the surrogate enables rapid evaluation of the full mechanical-property landscape, providing a $10^5$-fold acceleration relative to direct DFT evaluation.

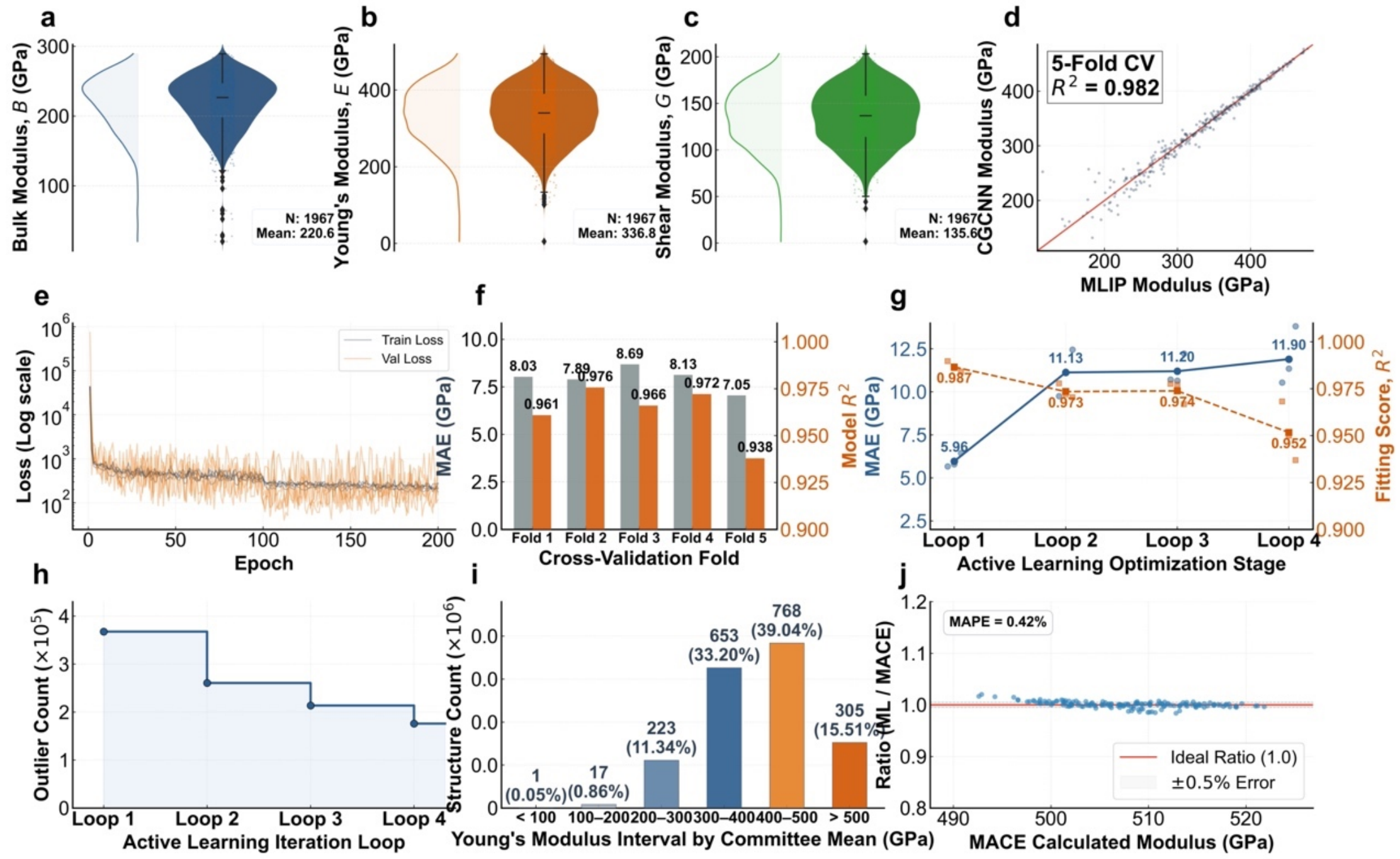

**Figure 4. Active-learning workflow and large-scale mechanical property screening.** **a**, **b and c** Modulus distributions ($B$, $E$, $G$) of the initial seed dataset. The 5-fold cross-validation performance of the CGCNN model, including **d** parity comparison, **e** loss evolution, and **f** fold-resolved error metrics. Active-learning convergence showing **g** MAE/$R^2$ trajectories and **h** elimination of high-uncertainty candidates across iteration loops. **i** Young's modulus distribution across the screened candidate space. **j** Direct MACE validation ratios (ML/MACE) within ±0.5 % error window.

## Mechanical Landscapes and Unconventional Thermal Transport at Extreme Temperatures

To further explore the mechanical-property landscape across the 7.7-million-configuration space, we evaluated the elemental co-occurrence statistics of Young's modulus (Fig. 5a). Pairings among refractory group-IV and group-V transition metals, including Ta-Ti, Ta-V and Hf-Ta, consistently exhibit high stiffness, with the Ta-Ti combination reaching 386.6 GPa compared with 360.1 GPa for Ta alone. Tracking Young's modulus as a function of solute concentration from 0 to 50 at.% reveals a distinct crossover near 12 at.% (~335 GPa; Fig. 5b). Below this concentration, the modulus varies relatively weakly with solute identity, whereas above it the elemental responses separate into distinct strengthening and softening regimes. Ta, Ti, V, Hf and Nb show the strongest stiffening, W and Zr produce moderate increases, Mo, Sc, Mn, Cr and Al have weaker effects, while Y and Ir produce pronounced softening. We further identified the 15 elemental combinations with the largest stoichiometric tunability, spanning Young's moduli from approximately 100 to 500 GPa (Fig. 5c). Notably, varying only the Ir/V ratio in the Cr-Ir-Ti-V-W-C system spans nearly the full mechanical range. Two configurations representing the high- and low-stiffness limits of this tunable system were therefore selected for detailed analysis: $Cr_2Ir_2Ti_{12}V_{22}W_2C_{40}$ with $E$=502.4 GPa and $Cr_{12}Ir_{20}Ti_2V_4W_2C_{40}$ with $E$ = 82.8 GPa. Before examining these limiting configurations, we assessed their finite-temperature vibrational response. Interaction-induced spectral broadening lowers the vibrational free energy in both systems, providing an additional entropic contribution to thermal stabilization (SI, Section 15 and Fig. S4). We then compared the calculated linewidths with the Ioffe-Regel criterion to assess the validity of the quasiparticle description (SI, Section 16 and Fig. S5). For the ultra-stiff $Cr_2Ir_2Ti_{12}V_{22}W_2C_{40}$ lattice, the linewidths remain below the Ioffe-Regel threshold up to 3600 K. In the compliant $Cr_{12}Ir_{20}Ti_2V_4W_2C_{40}$ lattice, mid-frequency modes cross the threshold above 1500 K, indicating the onset of an overdamped regime and limiting the quantitative applicability of the quasiparticle-based spectral free-energy treatment at higher temperatures.

We next examined whether the mechanically extreme configurations identified by the large-scale screening retain their stiffness at elevated temperatures. The top 50 quinary carbides with $E$ >500 GPa were selected for finite-temperature evaluation. Following the sampling protocol described above[33,34], perturbed supercells were evaluated with MACE and used to construct composition-specific Moment Tensor Potentials (MTPs)[35]. All 50 systems were dynamically stable, with harmonic phonon dispersions showing no imaginary frequencies (SI, Section 17 and Fig. S6). Finite-temperature MD simulations further showed that these carbides retain Young's moduli of 460-590 GPa up to 2100 K (Fig. 5d-f; Distillation of potential and MD simulation details are provided in SI, Section 18-20, Table S6 and Fig. S7). These values exceed those reported for conventional high-temperature structural alloys such as IN-718 and SS-304 (~150 GPa) and overlap or exceed the reported range for benchmark structural and thermal-barrier ceramics, including $\alpha$-$Al_2O_3$ and YSZ (~130-380 GPa)[36–40].

We finally examined the temperature dependence of lattice thermal transport. In temperature-dependent Ashby diagrams, these carbides occupy an unusual regime combining ultra-high elastic moduli with relatively low lattice thermal conductivity, deviating from the conventional stiffness-thermal-conductivity correlation (Fig. 5d-f). Decomposition of the heat current into particle-like and coherent wave-like channels clarifies this behavior (SI, Section 21 for detailed calculation). At 300 K, the two channels contribute comparably, with both $\kappa_p$ and $\kappa_c$ spanning approximately $20\ \mathrm{W\,m^{-1}\,K^{-1}}$. Upon heating to 1200 and 2100 K, anharmonic Umklapp scattering strongly suppresses the particle-like contribution to $0.5\text{-}3.0\ \mathrm{W\,m^{-1}\,K^{-1}}$, comparable to the lattice thermal conductivity of dense YSZ ($1.8\text{-}2.3\ \mathrm{W\,m^{-1}\,K^{-1}}$). In contrast, the coherent contribution remains nearly temperature-independent at $10\text{-}30\ \mathrm{W\,m^{-1}\,K^{-1}}$, becoming the dominant transport channel at high temperature (Fig. 5e,f). Thus, strong anharmonic suppression of particle-like transport can coexist with high elastic stiffness, producing a pronounced thermal-elastic decoupling in these multicomponent carbides.

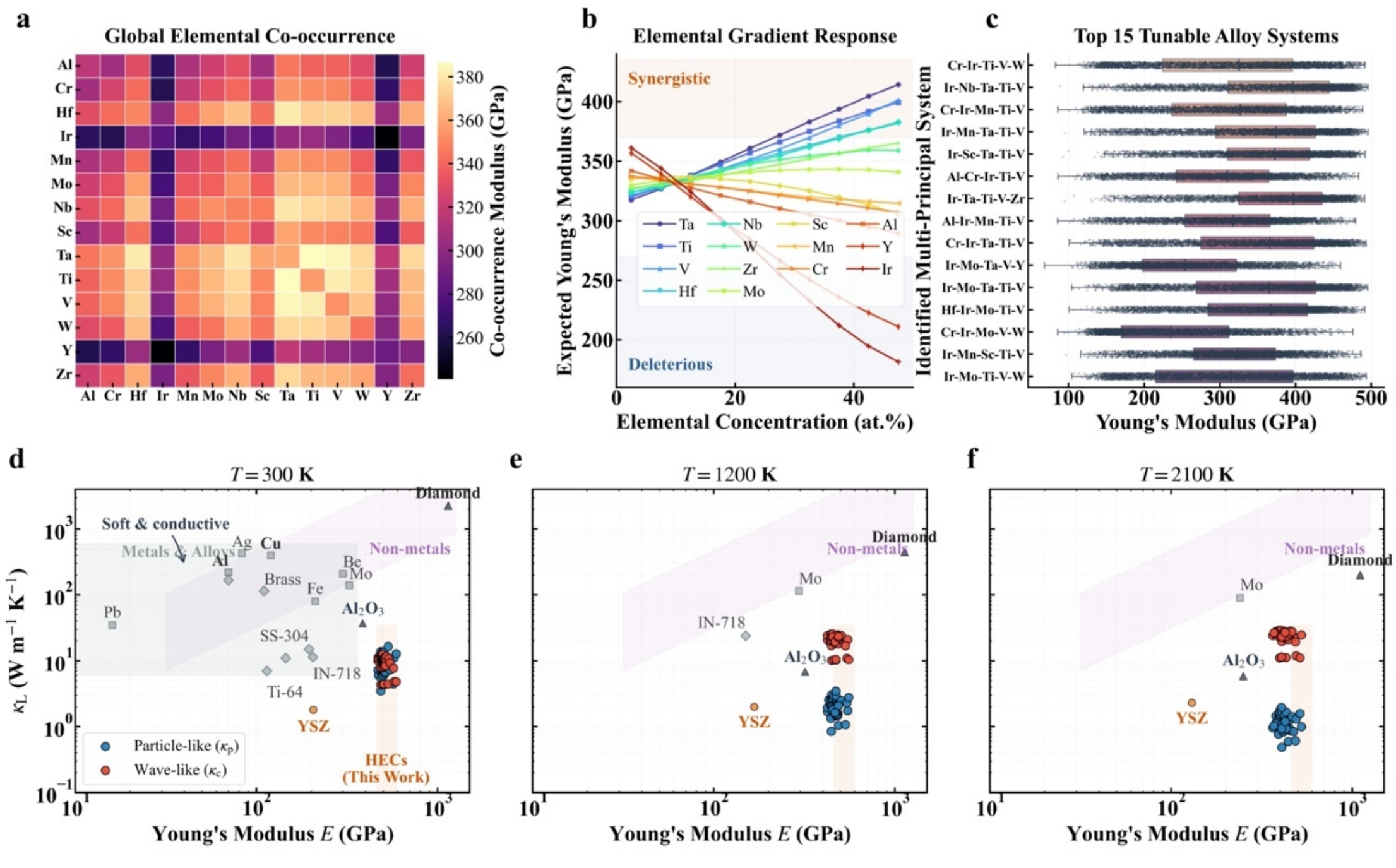

**Figure 5 | Compositional screening patterns and extreme mechanical-thermal property frontiers of high-entropy ceramics. a** Global co-occurrence expectation matrix of Young's modulus across $7.7 \times 10^6$ candidate compositions. **b** Elemental concentration trajectories of expected Young's modulus. **c** Modulus tuning spans across the top 15 identifiable elemental combinations. Ashby-style property maps at **d** 300 K, **e** 1200 K, and **f** 2100 K comparing the top 50 stiffest candidate configurations ($E > 500$

GPa at 0 K) with conventional structural alloys, benchmark ceramics ($Al_2O_3$, diamond), and thermal barrier oxides (YSZ). Young's moduli were obtained from molecular dynamics tensile simulations using MTP machine-learning potentials. Total lattice thermal conductivity is decomposed into particle-like propagation ($\kappa_p$, blue) and wave-like coherent ($\kappa_c$, red) contributions.

**Microscopic Origins of Mechanical and Thermal Responses**

To elucidate the microscopic mechanisms underlying the mechanical tunability and anomalous temperature dependence of lattice heat transport identified above, we performed real-space interatomic force-constant (IFC) decompositions and electronic-structure analyses for the two selected configurations above (Fig. 6). We first examined the microscopic origin of shear softening at high Ir concentrations by projecting the first-nearest-neighbors harmonic IFC matrices $\mathbf{\Phi}_{ij}$ onto the bond direction $\hat{\mathbf{r}}_{ij}$. The resulting longitudinal bond-stretching and transverse bond-shearing stiffnesses are defined as $K_L = \hat{\mathbf{r}}_{ij}^T \mathbf{\Phi}_{ij} \hat{\mathbf{r}}_{ij}$ and $K_T = \frac{1}{2}[\mathrm{Tr}(\mathbf{\Phi}_{ij}) - K_L]$, respectively (Fig. 6a,b). In the rigid $Cr_2Ir_2Ti_{12}V_{22}W_2C_{40}$ configuration, the lattice retains robust and relatively isotropic transverse restoring forces, with $K_T \approx 0.6\text{-}1.1 eV/Å^2$ and $K_L/K_T \approx$ 5-6 (Fig. 6a). By contrast, increasing the Ir concentration to 50% on the metal sublattice strongly suppresses the transverse restoring force: $K_T$ falls below 0.4 $eV/Å^2$, while $K_L/K_T$ increases to approximately 30-40, indicating pronounced shear softening (Fig. 6b).

Electronic and local-structural analyses reveal the origin of this loss of transverse restoring force (Fig. 6d,e). In the rigid configuration, the constituent metals retain near-$O_h$ coordination, while the charge-density difference shows symmetric charge accumulation along the metal-carbon bonds, consistent with a directional $\sigma$-bonding network (Fig. 6d). At 50% Ir occupancy on the metal sublattice, the increased chemical heterogeneity is accompanied by substantial local coordination distortion: Ir develops pronounced off-axis displacements, while neighboring Cr, Ti, V and W octahedra undergo cooperative tilting (Fig. 6e). These local distortions modify the directional M-C bonding environment and reduce the transverse restoring forces, providing a microscopic origin for the observed shear softening and macroscopic modulus reduction.

The lattice dynamics reveal the origin of the anomalous temperature dependence of thermal transport (Fig. 6c). In $Cr_2Ir_2Ti_{12}V_{22}W_2C_{40}$, the wave-like coherent thermal conductivity ($\kappa_c$) doubles from 0.57 $W\,m^{-1}\,K^{-1}$at 300 K to 1.27 $W\,m^{-1}\,K^{-1}$ at 1200 K, while the particle-like channel ($\kappa_p$) drops below 0.1 $W\,m^{-1}\,K^{-1}$. Frequency-resolved coherent spectral-density maps reveal how coherent transport in the chemically disordered lattice is redistributed across the phonon spectrum with increasing temperature (Fig. 6f,g). At 300 K, coherent transport is concentrated mainly in the low-frequency acoustic and sub-optical bands below 11.0 THz. Upon heating to 1200 K, the high-frequency carbon-dominated optical manifold at 14.5-23.0 THz becomes increasingly involved in coherent transport, extending the spectral weight across a broader frequency range. The increased contribution from these optical modes therefore coincides with the rise in $\kappa_c$ despite the increasing damping of the low-frequency acoustic modes.

The composition dependence of coherent transport is further resolved by the mode-resolved coherence times $\tau_{ij}$ (Fig. 6h). At 300 K, the Ir-rich $Cr_{12}Ir_{20}Ti_2V_4W_2C_{40}$ configuration exhibits a higher baseline coherent conductivity than $Cr_2Ir_2Ti_{12}V_{22}W_2C_{40}$, accompanied by a dominant coherence-time feature near 0.040 ps shifted into the low-frequency acoustic range of 0-5 THz. The shift of the dominant coherence-time feature towards the acoustic regime places a larger fraction of the coherent response in low-frequency modes, which possess relatively high group velocities and are thermally populated at room temperature. Together, these results show that chemical and mass disorder reshape the vibrational landscape across both mechanical and thermal responses, with local structural heterogeneity and mass fluctuations governing the frequency-dependent coherent transport response.

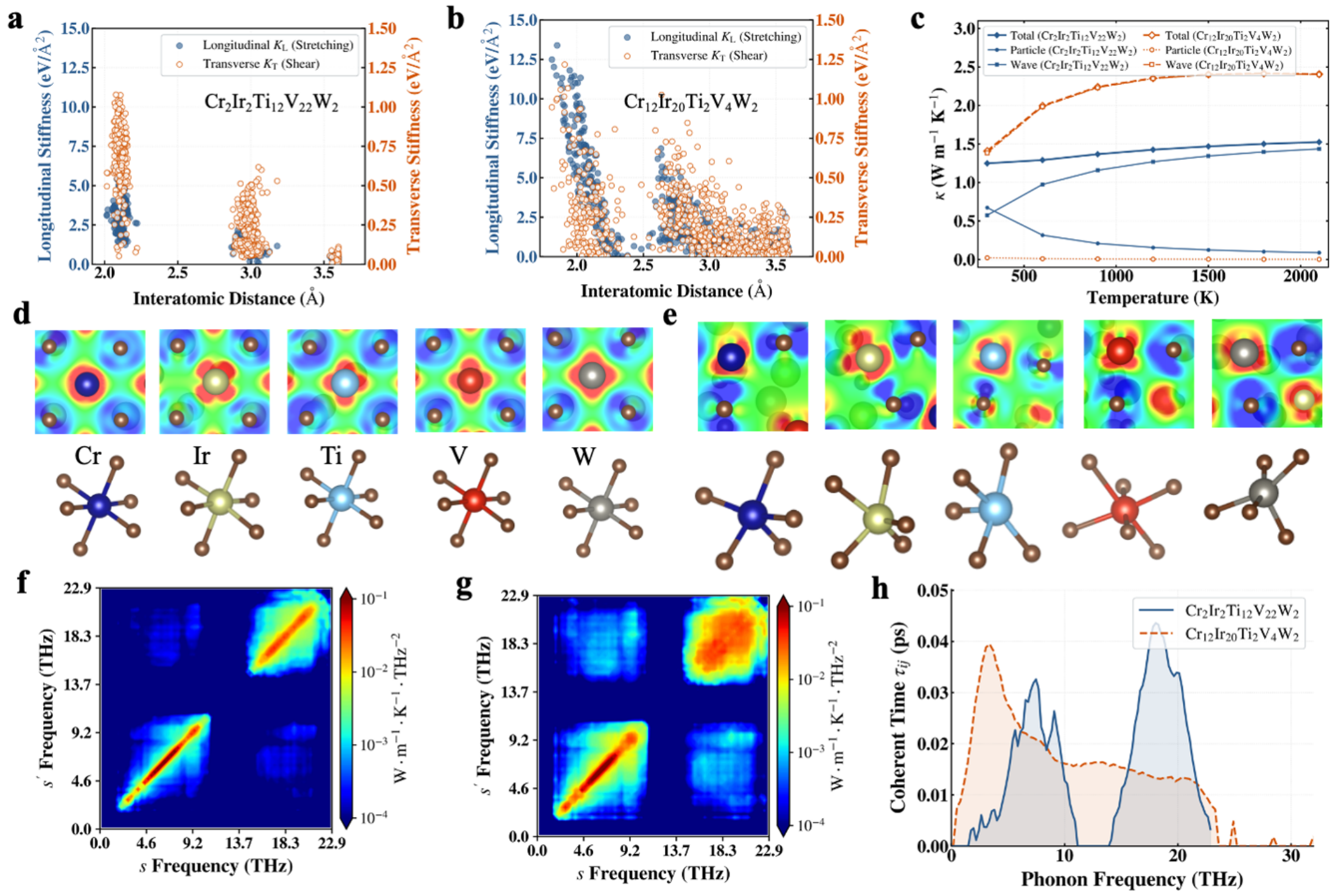

**Figure 6 | Microscopic Origins of Shear Softening and Anomalous Coherent Thermal Transport.** First-nearest-neighbor harmonic IFC profiles decomposed into longitudinal stretching and transverse shearing stiffnesses for the **a** rigid $Cr_2Ir_2Ti_{12}V_{22}W_2C_{40}$ and **b** compliant $Cr_{12}Ir_{20}Ti_2V_4W_2C_{40}$ configurations. **c** Temperature-dependent lattice thermal conductivity ($\kappa$) decomposed into particle-like and wave-like coherent channels for both phases. Local coordination polyhedra and charge density difference slices along the metal-carbon bonding planes for the **d** rigid and **e** compliant configurations. Two-dimensional frequency-resolved coherent spectral conductivity maps for $Cr_2Ir_2Ti_{12}V_{22}W_2C_{40}$ at **f** 300 K and **g** 1,200 K. **h** Comparison of mode-resolved coherent time scales between both configurations at 300 K. Carbon stoichiometry is omitted in panel labels for brevity.

## Conclusion

This study establishes a quantum many-body framework connecting local chemical bonding to high-temperature phase stability and macroscopic properties in multicomponent solids. Using finite-temperature many-body Green's function theory, we show that chemical disorder induces phonon spectral broadening through interaction-driven self-energy effects, lowering the vibrational free energy and providing an intrinsic entropic contribution beyond ideal configurational entropy that suppresses phase separation. Guided by active learning with universal interatomic potentials, we map 7.7 million disordered carbide configurations, revealing a distinct crossover near 12 at.% solute concentration that separates strengthening and softening regimes and enables broad mechanical property tuning. Microscopic analysis further connects shear softening to composition-dependent local coordination distortions, octahedral tilting and suppressed transverse restoring forces. Extended dynamical simulations reveal multicomponent lattices that retain high stiffness at elevated temperatures while sustaining relatively low lattice thermal conductivity and appreciable coherent heat transport. By bridging chemical bonding, quantum spectral thermodynamics and large-scale exploration of chemical space, this work provides a physically grounded framework for understanding and tuning the coupled thermodynamic, mechanical and thermal properties of chemically disordered solids.

## Acknowledgements

This work was supported by the Science and Technology Commission of Shanghai Municipality (Grant Nos. 25511103400 and 24CL2901702). The authors acknowledge the computational support provided by Shanghai Technical Service Center of Science and Engineering Computing, Shanghai University. We also thank Dr. Zeyu Wang for fruitful discussions on the machine learning methodology.

## Competing interests

The authors declare no competing interests.

## Data availability

The data supporting the findings of this study are available from the corresponding authors upon reasonable request.